\documentstyle[aps,multicol,prl,epsfig]{revtex}
\def\bea{\begin{eqnarray}}
\def\eea{\end{eqnarray}}
\def\be{\begin{equation}}
\def\ee{\end{equation}}
\begin{document}
\draft
\author{Zbyszek P. Karkuszewski, Krzysztof Sacha, and
Jakub Zakrzewski}
\address{
 Instytut Fizyki imienia Mariana Smoluchowskiego,
  Uniwersytet Jagiello\'nski,\\
 ulica Reymonta 4, PL-30-059 Krak\'ow
}
\title{A method for collective  excitation of Bose-Einstein condensate}
\date{\today}
\maketitle
\begin{abstract}
It is shown that by an appropriate
 modification of the trapping potential one may
create collective excitation in cold atom Bose-Einstein condensate.
 The proposed
method is complementary to earlier suggestions. It seems to be
feasible experimentally --- it requires only
a proper change in time of the potential in
atomic traps, as realized in laboratories already.
\end{abstract}
\begin{multicols}{2}

Spectacular experimental realizations of Bose-Einstein condensate
(BEC) in cooled and trapped atomic gases
 \cite{cornell95,ketterle95,bradley97}
stimulated intensive
investigations of possible modifications, control and manipulations of
this new state of matter. Here a macroscopic sample of atoms
is in a well defined quantumstate. Thus several
typically  quantum mechanical phenomena may now be investigated on a
macroscopic level.

As an example of manipulation of the condensate one may consider the
splitting of the condensate into two parts \cite{ketterle97}, well separated
in space and yet coherent with each other. The latter property may be
tested by
superimposing, at some later time, the two parts and observation of the
interference fringes \cite{ketterle97,zoller96,wallis97}.
Another example is the leakage of atoms from the
condensate that
may be used to prepare an ``atom laser'' \cite{zoller96a,ketterle97a}.

Another fascinating possibilities are revealed when one considers
possible collective excitations of the condensate. Several schemes
have been proposed  to create either solitary waves or
vortices in the condensate. Both these types of excitations are
the solutions of the time-dependent Gross-Pitayevsky equation (GPE)
as appropriate for the mean field, effective single
particle description of the gas of weakly interacting bosons in the
limit of vanishing temperature (for reviews see
 \cite{parkins98,dalfovo99}).
 In analogy to nonlinear optics
\cite{optics} one may consider bright solitons (bell shaped
structures propagating without dispersion), dark solitons (with
a node in the middle -- an analog of the first excited state in the
noninteracting particles picture) or the intermediate grey
solitons.

The early propositions for creation of solitions in BEC utilized
collisions between spacially separated condensates
 \cite{reinhardt97,scott98}.
Soon it was realized that less violent approaches are also possible.
Typical for atomic laser control -- resonant Raman excitation scheme
-- to excite vortex states has been proposed \cite{marzlin97}.
This approach relies on the resonance condition which is, however,
modified
during the transfer process due to the nonlinearity of GPE.
Another possibility which takes the nonlinearity fully into account
is the adiabatic scheme of \cite{dum98}. It
utilizes effectively internal atomic
transitions combined with appropriate states of the condensate for
a controlled laser induced adiabatic transfer, populating
 solitonic or vortex solutions
of GPE, depending on the details of the process.
 The latter approach seems more robust against
typical experimental uncertainties.
A yet different approach
produces a phase shift between two parts of the condensate -- such
a phase imprinting method, originally proposed in \cite{dobrek99},
has been actually utilized to create dark
  solitons both in cigar
shaped BEC \cite{burger99} and in the spherically symmetric condensate
  \cite{denschlag00}. The same method has been successfully applied to
create vortices \cite{matthews99}. The latter have been also
demonstrated experimentally using laser stirring approach
\cite{madison00}.

The aim of this communication is to propose yet another scheme for
effective collective excitation of the BEC. The method is in some
sense similar, in another sense opposite, to the adiabatic passage
of \cite{dum98}. In the approach of \cite{dum98} one slowly
tunes the laser frequency following adiabatically the levels. The
transfer of population between two internal atomic states is
accompanied by an appropriate change of the condensate wavefunction
into a dark soliton, two-soliton or vortex solution of the GPE.
In our proposition, discussed below, we consider a {\it single}
internal state and sweep the laser across the trap modifying
in this way
the trapping potential.

For explanation of the effect  we assume first that the
condensate consists of non-interacting particles. While such a condensate
is not realized in nature, it may provide a good starting point for an analysis
of weakly interacting Bose gas. We show later that the picture remains
valid for interacting particles by considering the numerical example
with attractive atom-atom interactions \cite{bradley97}.

\begin{figure}
\centering
{\epsfig{file=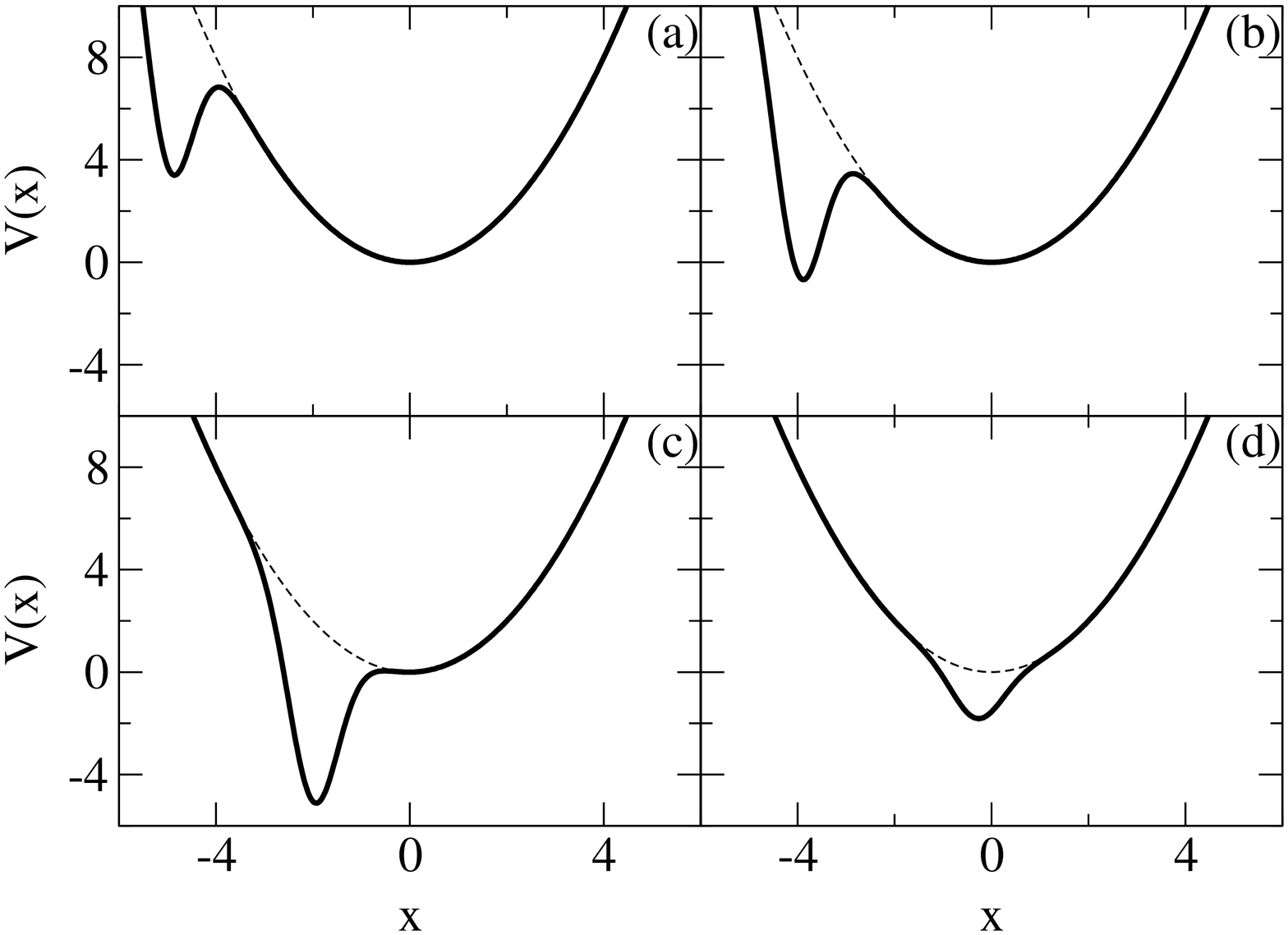, width=8.6cm}}
\caption{
Plots of the potential ({\protect \ref{poten}}) for $U_0=6.4$,
$\sigma=0.5$ and $x_0=-5.0$ (a), $x_0=-4.0$ (b), $x_0=-2.0$ (c) and $x_0=-0.3$
(d). Dashed lines give the unperturbed harmonic trapping well.
 }
\label{cont}
\end{figure}

To excite collectively a condensate we are going to modify trapping
harmonic potential along one of the independent directions only.
Therefore, as the non-interacting particles system is separable,
it is enough to consider atomic motion restricted to one
dimension (the generalization to a three dimensional case
is simple). Originally the condensate occupies the ground
state of the trap \cite{parkins98,dalfovo99,idziaszek99a}. 
Since we consider non-interacting particles it is
sufficient to consider a single particle picture. By imposing a
laser beam, being appropriately
tuned off (but close to) the resonance with respect to an internal
atomic transition we may modify the trapping potential by adding a
gaussian-shaped local well
\begin{equation}
V(x)=\frac{x^2}{2}+U_0\arctan(x_0)\exp\left(\frac{-(x-x_0)^2}{2\sigma^2}\right).
\label{poten}
\end{equation}
In the following, as above, we use the trapping harmonic oscillator units, i.e.
$\omega t$ for time and $\sqrt{\hbar/m\omega}$ for length, where $\omega$ is
harmonic oscillator frequency while $m$ stands for atomic mass.
Similar modification of the potential has been used to
split the condensate into two parts \cite{ketterle97} --- there instead of a
local well, a potential barrier has been created. We suggest here to
produce such a well on the very edge of the harmonic potential
(thus not affecting the condensate). Then we slowly sweep the well
across the potential (by moving the laser beam) simultaneously decreasing
the depth of the well (by adjusting the intensity of the beam) ---
it corresponds, for $U_0>0$, to a change of $x_0$ from some negative
value to zero, see Fig.~\ref{cont}.

\begin{figure}
\centering
{\epsfig{file=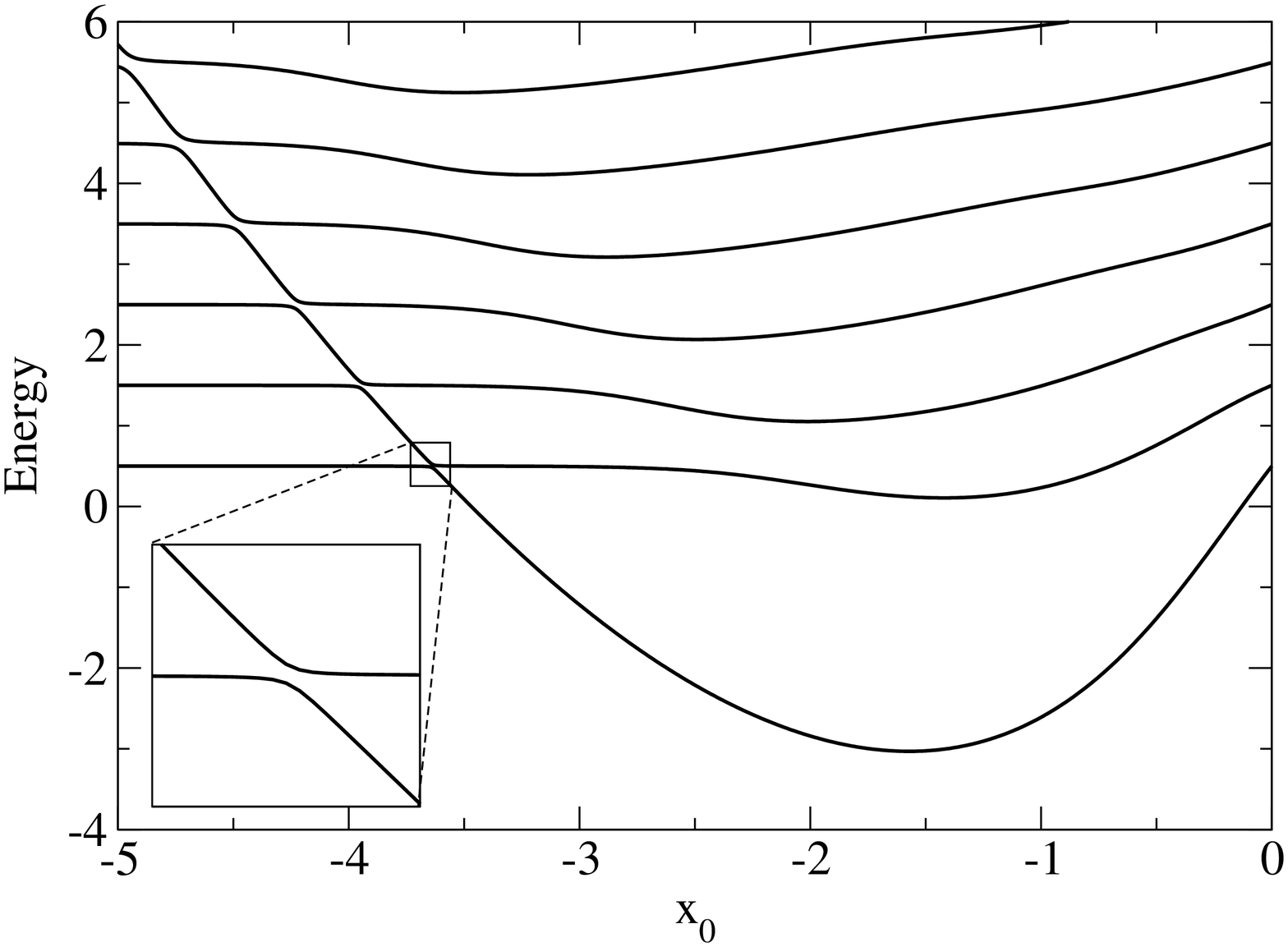, width=8.6cm}}
\caption{
Energy levels for a single particle in the potential
({\protect \ref{poten}}) for $U_0=6.4$ and $\sigma=0.5$ as a function of $x_0$.
Note the narrow avoided crossing between the ground and first excited states
around $x_0=-3.5$.
}
\label{ld}
\end{figure}

Assume that a particle is originally in the ground state of the
harmonic potential.
 For a sufficiently slow sweep the levels in
the ``time-dependent'' potential may be followed adiabatically
{\it except} in the vicinity of avoided crossings. By appropriately
choosing $U_0$ and $\sigma$ in Eq.~(\ref{poten}) we may arrange the situation
in which a narrow (with respect to a mean level spacing) avoided crossing
between
the ground and the first excited state of the potential occurs when
the local well sweeps the trap, see Fig.~\ref{ld}. If the avoided crossing is
narrow enough it may be passed {\it diabatically} and when the
local potential well disappears, the particle is left with a high
probability in the excited state. This is nothing else than the
Landau-Zener transition. The Landau-Zener effect has been explored in BEC
but for the transition of internal (not external) atomic degrees of
freedom \cite{ketterle97a}.

Assume that a single particle probability for the transition from the ground to
the lowest excited state is $p$.
Then for $N$ non-interacting particles originally being
in the ground state of the potential, the transition probability of
$k$ particles is $N\choose k$$p^k(1-p)^{N-k}$.
Thus mean energy of the condensate in the final state is
$(p+1/2)N\hbar\omega$ with variance $Np(1-p)\hbar^2\omega^2$.
For $N$ sufficiently large
(say of the order of thousands) and $p$ close to unity we get at
the end of the potential sweep a macroscopic part of the condensate
in the excited motional state. The final state of the
non-interacting condensate is
in fact describable as a time-dependent  wavepacket showing beats
with the frequency of the trapping potential. This is seen from the final
single particle reduced probability density which, independently on $N$,
reads
\begin{eqnarray}
|\Psi(x,t)|^2&=&(1-p)\psi_0^2(x)+p\psi_1^2(x) \cr
&& +2\sqrt{p(1-p)}\cos(\omega t)\psi_0(x)\psi_1(x),
\label{probden}
\end{eqnarray}
where $\psi_0(x)$ and $\psi_1(x)$ are harmonic oscillator ground and excited
states (in the real representation), respectively.

\begin{figure}
\centering
{\epsfig{file=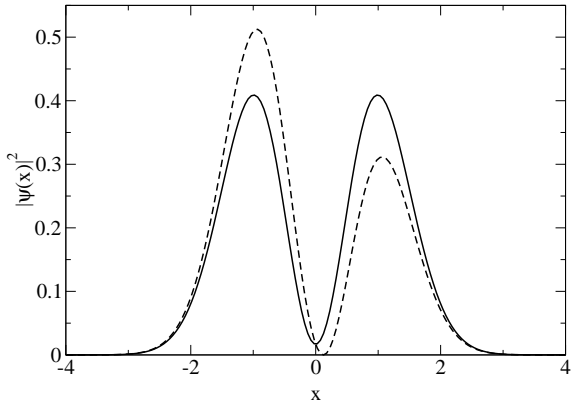, width=8.6cm}}
\caption{
Single particle reduced probability density of the condensate 
(corresponding to the noninteracting particles model), 
Eq.~({\protect \ref{probden}}), for $p=0.97$ at $\omega t=\pi/2$ (solid line) 
and $\omega t=\pi$ (dashed line).
}
\label{fig3}
\end{figure}

To check whether it is possible to realize an efficient transfer
using the method proposed we have simulated the situation
numerically. Choosing, without any special optimization attempt,
the parameters of the potential (\ref{poten}) as $U_0=6.4$, $\sigma=0.5$ and
changing $x_0$ from $-5$ to 0 with the velocity 0.1
we get $p=0.97$. The final single particle reduced probability density of the
condensate is then depicted in Fig.~\ref{fig3} at various times
of its periodic behavior.

As a specific example we propose in this communication to excite  
the condensate by sweeping the trapping
potential using the local potential well. However, the excitation
may be realized
in different ways --- the key point is to
arrange, in the level dynamics, a narrow isolated avoided crossing between
the ground and excited states.

One may argue that the proposed model of non-interacting particles
is very simple. The role of the interactions may be
subtle.
They clearly modify
the energy levels of the system. Such a modification will be felt
mostly close to avoided crossings that may be shifted and broadened.
Will this spoil completely the proposed scheme?
In our believe it will not, although an adjustment of the laser beam
intensity and other parameters may be necessary to optimize the
transfer of population. As a test of this assumption we consider
the excitation of the BEC with attractive atom-atom interaction as
realized for Li atoms \cite{bradley97}. While the one-dimensional approach
for interacting atoms is not exact (nonlinearity couples different
degrees of freedom) a one-dimensional approach based on GPE is often
used and may be justified for asymmetric traps
 \cite{garcia98,jackson98,muryshev99,fedichev99,busch00}.
 
\begin{figure}
\centering
{\epsfig{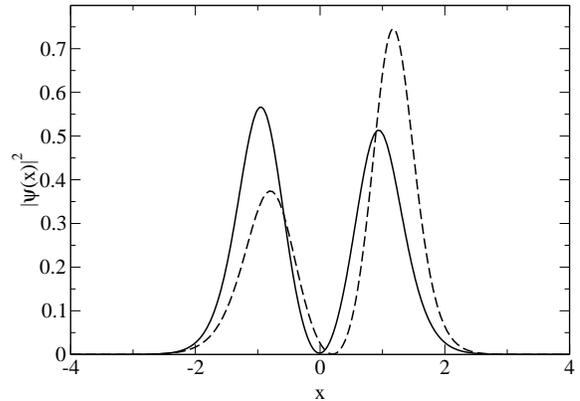}}
\caption{
The wavepacket obtained as a solution of the time-dependent GPE, 
Eq.~({\protect \ref{gpe}}), at the end of the potential sweeping (solid line)
and at the moment $\omega t=\pi$ later (dashed line).
}
\label{fig4}
\end{figure}

We integrate time-dependent GPE,
\begin{equation}
i\frac{\partial \psi}{\partial t}=-\frac{1}{2}\frac{\partial^2 \psi}{\partial x^2}
+V(x)\psi+g|\psi|^2\psi
\label{gpe}
\end{equation}
with $g=-5$ \cite{foot}, 
starting with the condensate in the ground state of the harmonic trap. 
By adjusting the parameters of
the potential (to compensate the influence of the interaction) taking
$U_0=10$, $\sigma=0.3$ and changing $x_0$ from $-5$ to 0 with the velocity 
0.05, we
were able to get a 97.5\% transfer of population into a
collective state corresponding to the first excited state
in the independent particle model. The final wavefunction obtained
via integration of the time-dependent GPE is depicted in
Fig.~\ref{fig4}.

It is interesting to compare the present ``diabatic'' approach
with the adiabatic scheme considered in \cite{dum98} (as the closest in
spirit among the techniques proposed).
 We must admit that the method of \cite{dum98} may be more robust and flexible,
in particular it is adaptable to excitation of vortices and multiple
solitons. The latter may be realized also by our method, one needs simply
to apply the sweeping potential twice or more times (this as well as
application to repulsive atom-atom interactions will be considered in
future). While the method of \cite{dum98} uses two internal states
(two component condensate) which is the common trend also in other
treatments of collective excitations, our approach considers a single
internal state. This may be advantageous in some applications.
Importantly also the adiabatic scheme \cite{dum98} takes necessarily
much longer time for an effective transfer (of the order of 200
or more periods of the harmonic trap) than our diabatic approach
(here a typical transfer time is 20 periods). While such
comparisons may be quite encouraging the best way
 of verifying our scheme would be
a laboratory test. Experimental setup requires only slight modifications
of the present atomic traps, thus, such an experiment can be realized
immediately.

To summarize we have proposed a simple scheme which enables us to create 
a collective excitation of the Bose-Einstein condensate. The proposed 
scheme may serve, we hope, as an alternative to other proposed and 
experimentally used already methods.

We are grateful to  Maciek Lewenstein and Kazik Rz\c{a}\.zewski for reading 
of the manuscript and several useful suggestions. Support of KBN under project 
No.~2P03B~00915 is acknowledged.


\end{multicols}
\end{document}